\documentclass[twocolumn,showpacs,preprintnumbers,amsmath,amssymb]{revtex4}

\usepackage{graphicx}

\begin{document}

\preprint{}

\title{Backaction noise produced via cavity-aided nondemolition measurement of an atomic clock state}

\author{Igor Teper, Geert Vrijsen, Jongmin Lee, and Mark A. Kasevich}

\affiliation {Physics Department, Stanford University, Stanford, CA
94305, USA}

\date{\today}

\begin{abstract}
We use a quantum nondemolition measurement to probe the collective
pseudospin of an atomic ensemble in a high-finesse optical cavity.
We analyze the backaction antisqueezing produced by the measurement
process to show that our protocol could create conditional spin
squeezing in the atomic ensemble.
\end{abstract}

\maketitle

    The application of recently developed techniques in cavity quantum
electrodynamics to many-atom ensembles is a promising means toward
the realization of novel experimental systems. The performance of
many protocols in quantum information processing \cite{Choi08,
Chaneliere05, Sherson06} that have been recently demonstrated in
free-space atomic systems is limited by the optical depth of the
atomic sample, which can be enhanced by placing atomic samples
inside optical cavities, as in \cite{Simon07}. Similarly, cavity
enhancement can facilitate the preparation of many-atom correlated
states, called squeezed states in analogy to squeezed light, that
can be used to surpass the usual projection noise limit for
atom-based sensors \cite{Kitagawa93, Wineland92}.

    Atomic spin-squeezed states have previously been produced by quantum
state transfer from squeezed light \cite{Hald99} or by entangling
two-photon transitions \cite{Meyer01, Fernholz08}, but much recent
work has focused on generating spin squeezing via quantum
nondemolition (QND) measurement \cite{Chaudhury06, Geremia05,
Oblak05, Kuzmich00}. A phase shift measurement protocol has been
used to entangle the collective spin states of two atomic ensembles
\cite{Julsgaard01}, and a QND scheme has been used to measure the
state of a superconducting charge qubit coupled to a microwave
resonator and to characterize the resulting measurement backaction
\cite{Schuster05}. A QND measurement using Rydberg atoms to
number-squeeze an intracavity photon state has also been
demonstrated \cite{Guerlin07}. Recently, the heating effects of
quantum-measurement backaction on an ultracold atomic gas have been
studied \cite{Murch08}.

    In this Letter, we report the implementation of a cavity-aided
continuous QND measurement of the pseudospin of a laser-cooled
$^{87}\rm Rb$ ensemble prepared in a superposition of
magnetic-field-insensitive hyperfine clock states. We present a
quantitative model for the interaction between the atoms and the
probe light and consider the predicted performance. We then describe
our experimental implementation, including considerations of
inhomogeneous coupling and the dephasing it causes, and provide
evidence, based on measurements of the backaction-induced
antisqueezing, that our protocol implements our model.

    An ensemble of $N$ two-level atoms coupled to a radiation field can
be described, in analogy with a spin-$\frac{1}2$ system, by a
collective pseudospin Bloch vector of length $J=N/2$, with $J_z$
corresponding to the population difference between the two states
\cite{Wineland92}. An off-resonant probe laser passing through such
an ensemble acquires a phase shift due to the atomic index of
refraction, which depends on $J_z$, without affecting $J_z$, so long
as spontaneous emission remains negligible. A subsequent observation
of the probe's phase projects $J_z$ onto the value that corresponds
to the observed phase shift. The fractional uncertainty in the probe
phase is limited by photon shot noise, which thus sets the limit on
the uncertainty in $J_z$ that can be far lower than the projection
noise limit for an uncorrelated state. In a cavity, the squeezing
factor, which measures the reduction in the projection variance, is
enhanced by a factor proportional to the square root of the cavity
finesse \cite{Sorensen02}, allowing much stronger squeezing than
possible in a free-space configuration.

    Quantitatively, assuming identical coupling to the
cavity for every atom and ignoring the spontaneous scattering of
probe photons into free space and the loss of photons through the
cavity mirrors, the Hamiltonian for the interaction of intracavity
probe light with the atomic state is given by (see, {\it e.g.},
\cite{Kuzmich98})

\begin{equation}
H=\hbar n g^2
\left[\frac{N}2\left(\frac{1}\Delta_2+\frac{1}\Delta_1\right)+J_z\left(\frac{1}\Delta_2-\frac{1}\Delta_1\right)\right],
\end{equation}

where $g$ is the atom-cavity coupling constant, $n=a^{\dagger} a$
and $N=J_{11}+J_{22}$ are the photon and atom number operators,
$J_z=\frac{1}2(J_{22}-J_{11})$ is the collective spin projection
operator, and $\Delta_{1,2}$ are the detunings from the lower and
upper clock states, respectively. Physically, the atoms experience
an AC Stark shift due to the presence of intracavity light, and the
light field experiences a phase shift due to the atomic index of
refraction. The term proportional to $N$ only adds an overall phase
shift to the light field, so only the term proportional to $J_z$ is
relevant for the clock performance analysis.

    In our model, the initial state of the system corresponds to a coherent atomic
state $|J,J_x\rangle$ and a coherent light state for the intracavity
probe field. The interaction leads to the imprinting on the probe
field of a phase shift $\Delta\phi=J_z\Omega t$ for
$\Omega=g^2(1/\Delta_{2}-1/\Delta_{1})$, where $t$ is the
interaction time, while the collective atomic pseudospin vector
precesses in the equatorial plane of the Bloch sphere at a rate
proportional to $n$. Since the probe beam's state is a superposition
of number eigenstates (Fock states), each of which causes the atomic
pseudospin vector to precess at a different rate, the uncertainty in
the atomic pseudospin vector's in-plane component ($J_y$ in the
frame that rotates with the atomic state so it remains polarized
along $x$) grows as it precesses. This growth in $J_y$ corresponds
to the quantum backaction of the measurement of its conjugate
variable $J_z$.

    Since the Hamiltonian entangles the collective atomic pseudospin
$J_z$ with the phase of the probe field, a measurement of the
probe's phase projects the atomic state onto a stochastically
determined state of $J_z$. For $n\Omega^2 t^2\ll 1$, the resulting
uncertainty in $J_z$, conditioned on the outcome of the phase
measurement, is given by $(\Delta J_z)^2=(N/4)(1+Nn\Omega^2
t^2/2)^{-1}$, compared to the usual projection noise of $(\Delta
J_z)^2=(N/4)$.

    For interaction times greater than the cavity photon lifetime, the above picture is
modified by the leakage of photons out of the cavity. We treat this
in a simplified way, by breaking up the interaction time into
intervals corresponding to the photon lifetime $\tau_{\text{cav}}$,
where, after each interval, the probe state is measured, and the
resulting atomic state then interacts with a new coherent photon
state. The repeated projection out of the intracavity photon state
destroys the atom-light coherence between interaction intervals,
causing the pseudospin variances to evolve linearly in time. The
squeezing for short times, until the antisqueezed uncertainty begins
to wrap around the Bloch sphere, is then given by $(\Delta
J_z)^2=(N/4)(1+N\bar{n}\Omega^2 t\tau_{\text{cav}}/\sqrt{2})^{-1}$,
where $\bar{n}$ now denotes the time-averaged intracavity photon
number, which depends not only on the input power but also on $t$
and $\tau_{\text{cav}}$ (in contrast to \cite{Nielsen08}, where all
the time dependence is included explicitly), and the corresponding
antisqueezing is $(\Delta J_y)^2=(N/4)(1+N\bar{n}\Omega^2
t\tau_{\text{cav}}/\sqrt{2})$.

    Atomic spontaneous emission into free space reduces the
correlation between the measured probe phase (to which the atoms
that have undergone spontaneous emission contribute) and the
collective pseudospin of the atoms used for the clock (to which they
don't contribute). If the spontaneous emission rate is known, the
average phase shift due to the atoms that have scattered can be
subtracted out, but the stochastic fluctuations in that phase shift,
given by the shot noise on the number of spontaneous emissions,
cannot be accounted for and will degrade the conditional squeezing
\cite{Bouchoule02}.

    Our experiment uses a hemispherical optical cavity, previously described in
\cite{Tuchman06}, with length $L=10$ cm, finesse $F=205000$, free
spectral range (FSR) of 1.505 GHz, HWHM linewidth
$\kappa/(2\pi)=3.7$ kHz, giving $\tau_{\text{cav}}=21.5$ $\mu$s, and
TEM$_{\rm 00}$ mode size of 310 $\mu$m at the atomic position,
corresponding to a maximum atom-cavity coupling $g/(2\pi)=53$ kHz
for the $|F=2, m_F=2 \rangle \rightarrow |F'=3, m_F=3 \rangle$
cycling transition, to probe the collective pseudospin state of a
cloud of $^{87}\rm Rb$ atoms cooled in a magneto-optical trap (MOT)
located at the center of the cavity. The atoms are released from the
MOT, further cooled by optical molasses, prepared in an equal
superposition of the $|F=1, m_F=0\rangle$ and $|F=2, m_F=0\rangle$
clock states, and probed by a standing wave of intracavity light.

\begin{figure}
\includegraphics[width=3.3in]{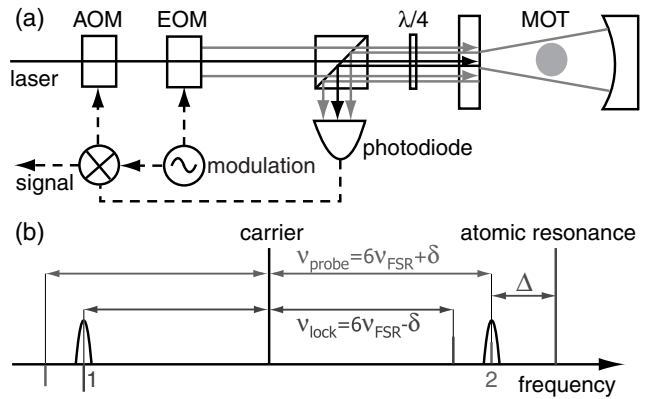}
\caption{Schematic of experiment (a) and modulation method (b). Two
sidebands resonant with the cavity are modulated onto a nonresonant
laser by an electro-optical modulator (EOM). A photodiode monitors
the beat notes between the carrier and the sidebands reflected from
the cavity. The demodulated beat note between the carrier and the
locking sideband (1) is used to drive a double-pass acousto-optical
modulator (AOM) that keeps the locking sideband resonant with the
cavity, while the beat note between the carrier and the probe
sideband (2) provides the experimental measurement of the probe
phase shift.}\label{fig1}
\end{figure}

    The modulation scheme we use to couple laser light into the cavity
and perform our measurements is a modified version of the one
described in \cite{Long07}, and is shown in Fig.~\ref{fig1}. We use
a far-detuned resonant sideband (``locking beam'') to stabilize the
laser to the cavity resonance via a Pound-Drever-Hall lock. The
collective atomic pseudospin is measured via the phase shift
produced by the atomic index of refraction on a near-detuned
resonant sideband (``probe''). The probe is tuned 1.5 GHz to the red
of the $5^{2}S_{1/2}, F=2 \rightarrow 5^{2}P_{3/2},F'=3$ atomic
transition with a cavity input power of 1.2-2.5 nW, while the
locking beam, 18.06 GHz farther to the red, has an input power of
2.5 nW.

    When we turn on the probe, the atom cloud has a 1/$e$ radius of 390 $\mu$m and is
falling through the cavity mode at 2.9 cm/s while expanding at 3.7
cm/s due to its temperature of 14.5 $\mu$K. While the atoms' passage
through the cavity limits the interrogation time to several ms, a
more stringent limit is imposed by inhomogeneous broadening due to
the presence of intracavity light. The intracavity standing waves
impose an AC Stark shift which varies depending on atom position
within the cavity mode, with an average of $U/k_{B}\sim$ 1 $\mu$K
(0.2 $\mu$K) for a 2.5 nW probe beam and $\sim$0.08 $\mu$K (0.06
$\mu$K) for a 2.5 nW locking beam, for atoms in the $|F=2,
m_F=0\rangle$ ($|F=1, m_F=0\rangle$) state. For the locking light,
which is always on, the 1/$e$ Rabi oscillation decay time for atoms
in the cavity is around 1 ms. For atoms prepared in an equal
superposition of the two clock states, which is most sensitive to
inhomogeneous broadening, the probe light dephases the collective
spin vector of the sample in several tens of $\mu$s.

    There are two distinct time scales for light-induced
inhomogeneous broadening, corresponding to the longitudinal and
transverse motions of the atoms in the cavity mode. The atoms are
unconfined by the light, and a typical atom's thermal velocity
causes it to travel the 390-nm distance between adjacent nodes in
the longitudinal standing wave in 10 $\mu$s, so measurements over
time scales longer than this should average over the longitudinal
inhomogeneities in the AC Stark shift. In the transverse direction,
however, the AC Stark shift varies by just a few percent in 100
$\mu$s of atomic motion, so its effects can be countered on that
timescale by using spin echo.

\begin{figure}
\includegraphics[width=3.3in]{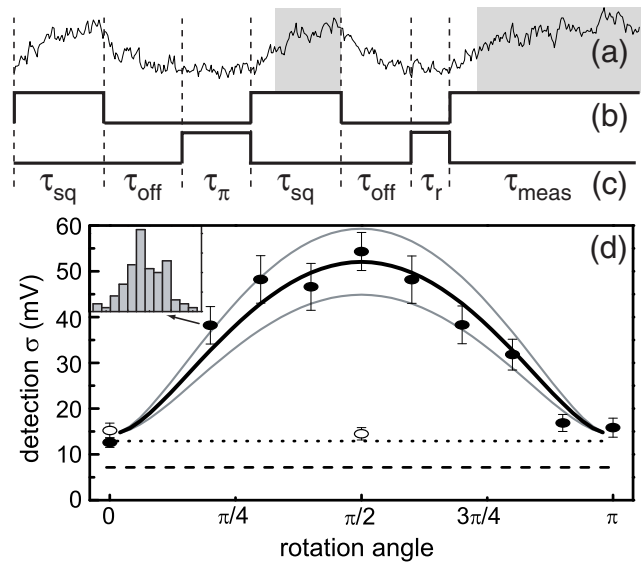}
\caption{Measurement protocol and results. The traces show a typical
probe signal (a) along with the control sequences for the probe
laser (b) and the microwave state rotation (c). The results obtained
for the fluctuations in the difference between the means of the two
shaded regions in (a), chosen to exclude the initial cavity buildup,
as we vary the duration $\tau_{\text{r}}$ of the final microwave
pulse are shown as solid circles in (d), with each data point
corresponding to 58-174 shots, with statistical error bars. The
inset in (d) shows a histograms of the difference of means for 87
shots at a final rotation of 0.63 radians, the width of which
corresponds to the indicated data point. The solid curve is a theory
calculation using our model, the gray curves account for
uncertainties in the experimental parameters, dominated by
uncertainty in intracavity probe power, the dashed line is the
expected projection noise, and the dotted line is the measured noise
floor in the absence of atoms (detector dark noise is negligible).
The open circles indicate the results obtained for inhomogeneous
broadening in the absence of spin echo (see text). The parameters
for this measurement are $\tau_{\text{sq}}=\tau_{\text {off}}=60 \;
\mu \text s$, $\tau_\pi=50 \; \mu \text s$, $N\approx57000$, probe
power $\approx$ 2.5 nW, locking power $\approx$ 2.5 nW.
}\label{fig2}
\end{figure}

    The correlated atomic state is generated and measured in a 3-pulse sequence
illustrated in Fig.~\ref{fig2}(a)-(c). The probe light is turned on
for a time $\tau_{\text {sq}}$, then turned off for a time
$\tau_{\text {off}}$ required for the probe light to leak out form
the cavity, after which a microwave $\pi$ pulse that lasts for a
time $\tau_\pi$ is used to prepare the spin echo. The probe light is
turned on for $\tau_{\text {sq}}$, which rephases the atomic spins,
then once again extinguished for $\tau_{\text {off}}$, and, after a
possible final microwave rotation, which takes a time $\tau_{\text
r}$, the probe light is turned on permanently and the atomic state
is measured for a time $\tau_{\text {meas}}$. During the two
preparation pulses, spontaneous emission for all data presented
here, measured by observing atom depumping over time, is
$\leq$6$\%$; for the final, destructive detection pulse, spontaneous
emission is 20-40$\%$. We use shot-to-shot fluctuations in the
difference in the cavity shift between the second squeezing pulse
and the final measurement to quantify the conditional projection
noise for our protocol.

    To properly calculate the expected projection noise, we must account
for spatially varying atom-cavity coupling. For a Gaussian atom
cloud with radius $r_a$ and a TEM$_{\rm 00}$ standing-wave cavity
mode with spot size $r_c$, the mean atom-cavity coupling $\bar
\Omega$ is less than the maximal on-axis, antinode value
$\Omega_{\text {max}}$ by a factor of $2((r_a/r_c)^2+1)$. The
variable that the probe phase measurement couples to is then not
$\Omega J_z$ but $\bar \Omega J_z,$ with an uncorrelated projection
noise variance given by the sum of the individual atom variances.
Since the inhomogeneous transverse coupling of different atoms does
not average out on the timescale of our measurement, this modified
collective variance scales not as the square of the mean coupling
but rather as the mean of the squared coupling: $(\Delta \bar \Omega
J_z)^2=\sum_{i=1}^N (\Delta \Omega_i j_{zi})^2=(N/16)\Omega_{\text
{max}}^2(2(r_a/r_c)^2+1)^{-1}$, where $i$ is an index over
individual atoms (the unmodified value is $(\Delta \Omega
J_z)^2=(N/4)\Omega_{\text {max}}^2$). To take advantage of the
squeezing produced by our protocol, subsequent measurements also
need to couple to $\bar \Omega J_z$, which suggests that
interferometer readout should be performed using the cavity shift
\cite{Kuzmich04}.

    By applying a microwave pulse before the final detection pulse,
it is possible to rotate the uncertainty ellipse around its center
and use the atomic shift to measure its width in an arbitrary
direction. A $\pi/2$ pulse rotates the antisqueezed ($J_y$)
component of the collective atomic spin onto a population difference
($J_z$), which results in maximal noise on the shift, while a
smaller rotation produces a correspondingly smaller effect. The
results of such a series of measurements are shown in
Fig.~\ref{fig2}(d).

    To characterize the coherence of our atomic states, we vary the phase of the final microwave
pulse while keeping its duration fixed at $\tau_{\pi/2}$ by applying
a phase offset to the microwave oscillator that generated the pulse.
We thus scan the collective pseudospin rotation axis in the
equatorial plane of the Bloch sphere, which produces an oscillation
in the atomic populations, which we read out via the mean values of
the final probe measurement. For the final state produced by our
measurement, the contrast in this oscillation is about 73$\%$ of the
full contrast obtained under the same circumstances in the absence
of probe squeezing pulses, which means that the length of the
collective spin vector on the Bloch sphere is reduced to 73$\%$ of
its initial length due to the relative dephasing of the individual
spins by the squeezing pulses, and the projection noise is reduced
by the same amount. We have also confirmed that the lock light has
no effect on atomic coherence by comparing the contrast measured
after the spin echo sequence without squeezing pulses to the initial
contrast measured immediately after the first $\pi/2$ pulse creates
the clock-state superposition.

\begin{figure}
\includegraphics[width=3.3in]{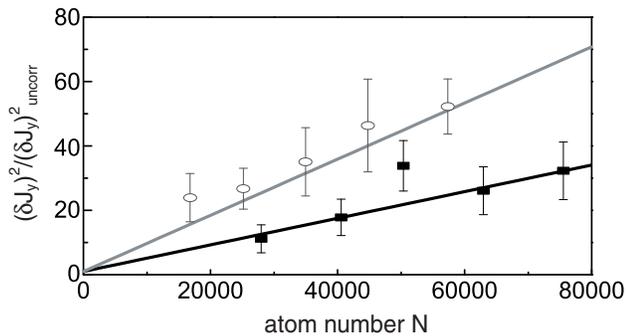}
\caption{Antisqueezing of the variance of the in-plane component of
the atomic pseudospin (conjugate of $J_z$) as a function of atom
number and probe intensity, obtained by rotating the spin state
after the two squeezing pulses (see Fig.~\ref{fig1}) by $\pi/2$ and
measuring the probe phase shift. The black squares and white circles
correspond to probe input powers of $\approx$ 1.2 nW and $\approx$
2.5 nW, respectively, with statistical error bars. The lines
correspond to theory calculations using our model; there is an
overall scaling uncertainty of $\pm30\%$ from probe power
measurements. Fitting the slopes for the two data sets as free
parameters gives $9.7(4)\times10^{-4}$ and $4.5(5)\times10^{-4}$,
with a ratio of 2.2(3), in good agreement with the ratio of the
input powers (2.1).}\label{fig3}
\end{figure}

    It is important to distinguish between inhomogeneous
broadening due to the spatially varying probe light intensity, which
leads to dephasing between the pseudospin states of different atoms,
and the dephasing of the collective pseudospin, i.e. antisqueezing.
To study the two effects independently, we use the fact that spin
echo counteracts the majority of the inhomogeneous dephasing, making
it possible to produce, without spin echo, the same amount of
inhomogeneous dephasing contrast reduction with a much shorter probe
pulse (and, consequently, much less antisqueezing). We find that a
20-$\mu \text s$ probe pulse without spin echo produces the same
amount of inhomogeneous dephasing (as measured by microwave
oscillation contrast) as two sequential 60-$\mu$s probe pulses with
spin echo. However, the 20-$\mu$s pulse does not produce a
measurable increase in noise (see open circles in
Fig.~\ref{fig2}(d)).

    We measured the antisqueezing as a function of atom number for two
different probe intensities. The results, along with theoretical
calculations, are shown in Fig.~\ref{fig3}. The observed linear
scaling with the atom number $N$ and with the probe intensity, and
therefore the average photon number $\bar{n}$, confirms the validity
of our measurement protocol. Variance due to classical intracavity
intensity fluctuations would scale quadratically with $\bar{n}$ (we
independently measure these fluctuations to be below 0.08$\%$
between the two squeezing pulses).

    While we are not able to observe squeezing for 60-$\mu$s squeezing
pulses, using the same protocol with longer pulses, which destroy
Rabi oscillation coherence, allows us to resolve $J_z$ to 3.8 dB in
the variance below the projection noise for an uncorrelated state
with spontaneous emission loss of $\leq$30$\%$. Technical
measurement noise and residual variations in atom-cavity coupling
limit our ability to observe squeezing while preserving coherence.
If the probe could be measured at the photon shot noise limit for
the second squeezing pulse, and taking into account the atomic loss
dictated only by the spontaneous emission and not by inhomogeneous
broadening, the antisqueezing shown in Fig. 3 would, for a
minimum-uncertainty state, correspond to spin squeezing of up to 10
dB in the variance. Significantly better performance could be
achieved by increasing the atom number (a typical $^{87}$Rb MOT with
$10^8$ atoms should give us measurement samples of $10^7$ atoms, a
hundredfold increase) and by improving the spatial overlap between
the cavity mode and the atom cloud, by either confining or cooling
the atoms.

    Our two-squeezing-pulse scheme has an implicit atom number measurement
(required for a practical clock or interferometer), since $J_z$
before the $\pi$ pulse becomes $-J_z$ after, so averaging the
measurements from the two pulses gives a phase shift that depends
only on the the total atom number. Combined with experimental
improvements that allow direct measurement of squeezing, our
protocol should significantly reduce the projection noise for atomic
metrology.

    We acknowledge funding support from DARPA and the MURI on Quantum
Metrology sponsored by ONR.

\end{document}